\documentclass[a4paper,11pt]{article}
\usepackage{jcappub} % for details on the use of the package, please
\usepackage{graphicx}
\usepackage{subfigure}
\usepackage{epstopdf}
\usepackage{comment}

\newcommand{\beq}{\begin{equation}}
\newcommand{\beqa}{\begin{eqnarray}}
\newcommand{\eeq}{\end{equation}}
\newcommand{\eeqa}{\end{eqnarray}}

\newcommand{\simgt}{\lower.5ex\hbox{$\; \buildrel > \over \sim \;$}}
\newcommand{\simlt}{\lower.5ex\hbox{$\; \buildrel < \over \sim \;$}}

\title{\boldmath Structure formation in a mixed dark matter model
with decaying sterile neutrino: the 3.5 keV X-ray line 
and the Galactic substructure}

\author[a]{Akira Harada,}
\author[b,c]{Ayuki Kamada,}
\affiliation[a]{Department of Physics, University of Tokyo, Tokyo 113-0033, Japan}
\affiliation[b]{Kavli IPMU (WPI), University of Tokyo, Chiba 277-8583, Japan}
\affiliation[c]{Department of Physics and Astronomy, University of California, Riverside, CA, 92507, USA}

\emailAdd{harada@utap.phys.s.u-tokyo.ac.jp}
\emailAdd{ayuki.kamada@ucr.edu}
\abstract{
We perform a set of cosmological simulations of structure formation in a mixed dark matter (MDM) model. 
Our model is motivated by the recently identified $3.5\,{\rm keV}$ X-ray line, which can be explained by the decay of non-resonantly produced sterile neutrinos accounting for $20\text{--}60\%$ of the dark matter in the Universe. 
These non-resonantly produced sterile neutrinos have a sizable free-streaming length and hence behave effectively as warm dark matter (WDM). 
Assuming the rest of dark matter is composed of some cold dark matter (CDM) particles, we follow the coevolution of a mixed WDM plus CDM cosmology. 
Specifically, we consider the models with the warm component fraction of $r_{\rm warm}=0.25$ and $0.50$.
Our MDM models predict that the comoving Jeans length at the matter-radiation equality is close to that of the thermally produced warm dark matter model with particle mass $m_{\rm WDM}=2.4\,{\rm keV}$, but the suppression in the fluctuation power spectrum is weaker.  
We perform large $N$-body simulations to study the structure of non-linear dark halos in the MDM models. 
The abundance of substructure is significantly reduced in the MDM models, and hence the so-called small-scale crisis is mitigated. 
The cumulative maximum circular velocity function (CVF) of at least one halo in the MDM models is in good agreement with 
the CVFs of the observed satellites in the Milky Way and the Andromeda galaxy.
We argue that the MDM models open an interesting possibility to reconcile the reported $3.5\,{\rm keV}$ line \emph{and} the internal structure of galaxies.}


\begin{document}

\maketitle
\flushbottom

\section{Introduction}
\label{sec:intro}
It has been well established that the standard Lambda cold dark matter ($\Lambda$CDM) model is consistent with an array of observations of large scale structure, such as galaxy clustering and anisotropies of the cosmic microwave background (CMB)\,\cite{2014MNRAS.440.2692S, 2013ApJS..208...19H, 2014A&A...571A..16P}.
However, there appears to be inconsistencies between predictions of the $\Lambda$CDM model and observations at subgalactic scales ($\sim 10^{8\text{--}10}\,M_\odot$ objects), which is often called the \emph{small-scale crisis}.
One of the problem is the \emph{missing satellite problem}\,\cite{Moore1999, Kravtsov2010}, which claims that the cumulative maximum circular velocity function (CVF) of subhalos 
\footnote{Locally overdense and self-bound dark matter (DM) clumps 
that reside in the DM halo.} 
in the $\Lambda$CDM simulation does not match that of the Local Group. 
Not only within the Local Group, but the subgalactic structure in field galaxies probed by the $\rm H_I$ disk velocity width function also suggests essentially the same problem\,\cite{Zavala2009, Papastergis2011}.
A promising solution to the small-scale crisis is to introduce warm dark matter (WDM)\,\cite{Colin2000}. 
Free-streaming of WDM particles effectively suppresses the growth of small-scale structures and reduces the abundance of subhalos. 
According to popular particle-physics models, the degree of particle free-streaming is inherently related to the particle
mass.
One can place an upper bound on the WDM particle mass from observations of the abundance and the distribution of 
Galactic satellite galaxies and those of field low-mass galaxies if baryonic effects are ignored\,\cite{Schneider2014}.
\footnote{Baryonic effects can also be important in solving the small-scale crisis. 
We provide further discussion of baryonic effects in section\,\ref{sec:conclusion}. 
The observations of the Galactic satellites can also be used to put a conservative lower bound on the WDM particle mass if the baryonic effects are ignored \,\cite{Polisensky2011, Lovell2014, 2014PhRvD..89b5017H}, while the bound is weaker compared with that from the observations of the Lyman-$\alpha$ forests.}
Interestingly, observations of the Lyman-$\alpha$ forests in distant quasar spectra provides lower bounds on the DM particle mass from the measurement of small-scale clustering of neutral $\rm H_I$ gas\,\cite{Viel2013}. 
Unfortunately, these constraints actually contradict each other\,\cite{Schneider2014}, suggesting that introducing a pure WDM does not provide a consistent solution to the small-scale crisis. 
It thus appears important and timely to explore another model that can be made consistent with both the observations.
In the present paper, we study structure formation in mixed dark matter (MDM) models 
\footnote{\label{fn:mdm} In the present paper, we use the term ``mixed dark matter''\,\cite{2012JCAP...10..047A} to represent cold + warm dark matter, which is often called CWDM model in the literature.} 
that can relax the constraint from the Lyman-$\alpha$ forests\,\cite{Boyarsky2009}.

From the viewpoint of DM indirect search, there exists another motivation for the MDM model. 
Two recent X-ray observations\,\cite{Bulbul2014, Boyarsky2014} independently indicate the existence of an unidentified X-ray line at $E_\gamma \simeq 3.5\, {\rm keV}$ in stacked spectra of galaxy clusters (e.g., the Perseus galaxy cluster) and the Andromeda galaxy observed by \emph{XMM-Newton} and \emph{Chandra}.
\footnote{An unidentified X-ray line is also reported at $E_\gamma \simeq 2.5\, {\rm keV}$ in \emph{Chandra} observation of Willman 1\,\cite{2010ApJ...714..652L}.
This $2.5\, {\rm keV}$ line is not confirmed but not ruled out in \emph{XMM-Newton} follow-up observation\,\cite{2012ApJ...751...82L}.
For current instruments, however, it is challenging to distinguish unidentified line features from atomic lines or instrumental effects.
Higher resolution spectroscopy in, for instance, coming \emph{ASTRO-H} mission\,\cite{2012SPIE.8443E..1ZT} is expected to confirm or reject their dark matter origin.} 
The unidentified line can be originated from the elementary process of DM particles. 
Many studies attempt to explain the unidentified line in particle-physics motivated DM models\,\cite{2014PhLB..732..196I, 2014arXiv1402.6671F, 2014PhLB..733...25H, 2014PhRvD..89j3511J, 2014arXiv1403.0865L, Abazajian2014, 2015arXiv151204544B, 2015arXiv151204548H, 2014JCAP...09..007K, 2014arXiv1403.1280E, 2014arXiv1403.1536K, 2014JCAP...05..033F, 2014arXiv1403.1710B, 2014PhLB..735..338N, 2014PhLB..735...92C, 2014PhRvD..90b3540C, 2014PhRvD..90a3007T, 2014PhRvD..90b3535K, 2014PhRvD..89l7305A, 2014JHEP...06..037D, 2014PhRvD..90a1701B, 2014JCAP...05..044P, 2014PhLB..734..178N, 2014arXiv1403.7742K, 2014PhRvD..90c5014D, 2014PhLB..735...69Q, 2014PhRvD..90c5002D, 2014PhRvD..89k5011B, 2014arXiv1404.3676P, 2014PhRvD..89l1302C, 2014arXiv1404.4795O, 2014arXiv1404.5446L, 2014arXiv1404.7118R, 2014arXiv1404.7741C, 2014arXiv1405.3730B, 2014arXiv1405.4670N, 2014arXiv1405.6967C, 2014arXiv1406.0660K, 2014PhRvD..90c5009C, 2014arXiv1406.5518C, 2014arXiv1406.5808I, 2014arXiv1406.6481G, 2014arXiv1406.6556A, 2014arXiv1407.0460C, 2014arXiv1407.0863D, 2014arXiv1407.2543O, 2014arXiv1407.6827H, 2014arXiv1408.0233C, 2014arXiv1408.0286H, 2014arXiv1408.2950F, 2014arXiv1408.3936H, 2014PhLB..738..380F}. 
Here, we focus on the case that the radiative decay of the sterile neutrino with mass $M \simeq 7 \,{\rm keV}$ is the origin of the X-ray anomaly. Furthermore, we consider that the sterile neutrino is produced via non-resonant process (Dodelson-Widrow mechanism\,\cite{Dodelson1994}) in the early Universe. 
The resonantly produced sterile neutrino is revisited in response to the X-ray anomaly in ref.\,\cite{Abazajian2014, 2015arXiv151204544B, 2015arXiv151204548H}. 
The resonant production requires late time production of large lepton asymmetry, while the non-resonant production does not. 
It is important to investigate if the non-resonantly produced (NRP) sterile neutrino can explain the X-ray anomaly and how it affects small-scale structures.
With the above assumptions, our model predicts that the amount of NRP sterile neutrino is less than the total amount of DM in the Universe\,\cite{2013ApJS..208...19H}. 
We thus assume that another stable particle accounts for the rest of DM that effectively behaves as CDM. 
It is worth mentioning that such a multi-component dark matter model opens up an interesting possibility. 
The rest of DM can be composed of weakly interacting particles (WIMPs).
Their annihilation can explain the reported $\gamma$-ray anomaly in the Galactic center\,\cite{2011PhRvD..84l3005H, 2013PhRvD..88h3521G, 2014arXiv1402.4090A, 2014arXiv1403.1987L}, especially if their thermal relic contributes only partially to the total dark matter density. 
The sterile neutrino that is produced non-resonantly in the early Universe has sizable free-streaming length and behaves as the WDM particle. 
The observations of the Lyman-$\alpha$ forests\,\cite{Boyarsky2009} suggest that the NRP sterile neutrino with $M \simeq 7\,{\rm keV}$ is excluded if it accounts for the whole DM. 
MDM models can evade this constraint, however. 
It is thus important and timely to study structure formation in MDM models, especially from the viewpoint of the small-scale crisis. 
\footnote{The statistical properties of field halos are addressed in MDM models with different parameter sets in ref.\,\cite{2013MNRAS.428..882M}.}

In the present paper, we first review the particle-physics motivation and the production mechanisms of the sterile neutrino. 
Then we identify the parameter region in which the recently detected X-ray line can be explained. 
This is presented in section\,\ref{sec:neutrino}. 
We show the results of linear and non-linear structure formation in MDM models with the identified parameters in section\,\ref{sec:evolution}. 
Finally, in section\,\ref{sec:conclusion}, we give discussion and concluding remarks.

\section{Sterile neutrino}
\label{sec:neutrino}
This section is aimed at showing that when the NRP sterile neutrino occupies a certain fraction of the whole DM mass density, its radiative decay can explain the observed X-ray anomaly. 
We begin with a brief summary of the particle-physics motivation and the properties (especially, mixing angle) of the sterile neutrino.
Then, we describe in detail the production mechanism and derive the resultant relic density, which determines the NRP sterile neutrino fraction.
The calculation of the relic density can be found in previous works \cite{Dodelson1994, 2002APh....16..339D, 2001PhRvD..64b3501A, 2002PhRvD..66b3526A, 2006JHEP...06..053A, 2007JHEP...01..091A}.
By following them, we clarify the difference between the work for the resonantly produced sterile neutrino in ref.\,\cite{Abazajian2014, 2015arXiv151204544B, 2015arXiv151204548H} and this work.
Furthermore, we show that the relic NRP sterile neutrino obey the Fermi-Dirac distribution up to a prefactor. 
The distribution function is incorporated in the calculation of the matter power spectra in the next section.
Finally, we specify the required NRP sterile neutrino fraction and the corresponding mixing angle to explain the observed X-ray anomaly.
After we specify the particle-physics model, we investigate structure formation in MDM models with the identified NRP sterile neutrino fraction in the rest of the paper.

Observed oscillations between different species of the left-handed neutrinos indicate that there should be a right-handed counterpart.
The existence of a right-handed counterpart, however, has not been established directly in existing experiments. 
In fact, it does not have to be the case as we see in the following.
In a simple extension of the Standard Model (SM), the right-handed neutrino $\nu_{{\rm R}, \, I}$ couples with the left-handed lepton doublet $L_{\alpha}$ and the SM Higgs doublet $H$ through a Yukawa coupling 
${\cal L}_{\rm int} = - \sum_{\alpha, \, I} F_{\alpha \, I} {\bar L}_{\alpha} H  \nu_{{\rm R}, \, I} + {\rm h.c.}$. 
Once the Higgs doublet develops a vacuum expectation value, $\langle H \rangle$=$(0,v)$, the Yukawa coupling gives a Dirac mass to the neutrino. 
To be gauge invariant, the Yukawa coupling requires the right-handed neutrino to be a singlet under the SM gauge group.
The inertness makes the right-handed neutrino inaccessible to existing experiments.
This is why the right-handed-like mass eigenstate is referred to as the sterile neutrino, whereas the left-handed-like mass eigenstate is referred to as the active neutrino.

Gauge invariance allows the right-handed neutrino to have a Majorana mass 
${\cal L}_{\rm int} = - \sum_{I} M_{I}/2 \, {\bar \nu}_{{\rm R}, \, I}^{c} \nu_{{\rm R}, \, I} + {\rm h.c.}$. 
Here we take a basis of the right-handed neutrino such that the Majorana mass matrix is diagonal.
The Majorana mass may lead to a hierarchy between the active and the sterile neutrino mass. 
We can find the mass eigenstate $\nu^{\rm m}_{i}$ by diagonalizing the mass matrix.
When an unitary matrix $U$ relates these two bases such that 
$\nu^{\rm m}_{i} = \sum_{\alpha} U_{\alpha i} \nu^{c}_{{\rm L} ,\, \alpha} + \sum_{I} U_{I i} \nu_{{\rm R}, \, I}$,
the mixing angle $\theta$ is defined by $\sin^{2}(2 \theta) = 4 |\sum_{\alpha} U_{\alpha {\rm DM}} U^{*}_{\alpha i}|^{2}$.
The mixing angle is important for the phenomenology of the sterile neutrino since the sterile neutrino can interact with the SM particles only through the mixing. 
The reaction rate with SM particles is suppressed at least by a factor of $\sin^{2}(\theta)$.

Now let us discuss the production of the sterile neutrino in the early Universe.
In the present paper, we consider small mixing angles of $\sin^{2}(2 \theta) < 10^{-8}$. 
With such a small mixing, the sterile neutrino can not be thermalized in the plasma of the SM particles. 
The sterile neutrino is produced non-thermally via oscillations between the active and the sterile neutrino.
\footnote{Hereafter we consider the minimal framework of sterile neutrino introduced above. 
In possible extensions, sterile neutrinos can be produced in their intrinsic mechanisms as well as the active-sterile oscillation (see ref.\,\cite{2009PhR...481....1K} for review).
If one gauges $U(1)_{\rm B-L}$, for instance, right-handed neutrinos can be produced through exchange of the $U(1)_{\rm B-L}$ gauge boson immediately after the reheating of the Universe\,\cite{2010PhLB..693..144K}.
In another extension, one introduces a singlet Higgs boson. 
After the singlet Higgs boson develops a vacuum expectation value, the Majorana mass $M_{I}$ arises from Yukawa coupling to the right-handed neutrinos.
Non-thermal decay of the singlet Higgs boson can contribute to the relic density of sterile neutrino\,\cite{2006PhRvL..97x1301K, 2008PhRvD..77f5014P, 2008PhRvD..77j5004P}.} 
It should be noted that the oscillation is affected by the existence of SM particles especially in the early Universe\,\cite{1988NuPhB.307..924N}.
For simplicity, we focus on the case that only one active-sterile pair is relevant. 
Then, the Hamiltonian is given by
\begin{eqnarray}
{\mathcal H} \simeq {\rm diag}\{p+V_{\alpha}, p\} + \frac{1}{2p}
\begin{pmatrix}
0 & F_{\alpha} v \\
F_{\alpha} v & M 
\end{pmatrix}^{2}
\,,
\end{eqnarray}
where $p$ is the absolute value of the neutrino $3$-momentum, $V_{\alpha}$ is the potential for the left-handed neutrino, and $F_{\alpha}$ and $M$ are set to be positive by the redefinition of the neutrino states. 
Here we have assumed the neutrinos are relativistic, $p \gg F_{\alpha} v, M$.
After diagonalizing the Hamiltonian, we obtain the effective mixing angle,
\begin{eqnarray}
\sin^{2}(2\theta_{\rm eff}) &=& \frac{4 F_{\alpha}^{2} v^{2} M^{2}}{4 F_{\alpha}^{2} v^{2} M^{2} + (M^{2} - 2p V_{\alpha})^{2}} \\
&=& \frac{(\Delta m^{2}/2p)^{2} \sin^{2}(2\theta)}{(\Delta m^{2}/2p)^{2} \sin^{2}(2\theta) + \left\{ (\Delta m^{2}/2p) \cos(2\theta) - V_{\alpha}\right\}^{2}} \,, \label{eq:effectiveangle}
\end{eqnarray}
and the effective energy squared difference,
\begin{eqnarray}
\Delta m_{\rm eff}^{2}/2p &=& \sqrt{4 F_{\alpha}^{2} v^{2} M^{2} + (M^{2} - 2p V_{\alpha})^{2}} \\
&=& \sqrt{(\Delta m^{2}/2p)^{2} \sin^{2}(2\theta) + \left\{ (\Delta m^{2}/2p) \cos(2\theta) - V_{\alpha}\right\}^{2}} \,.
\end{eqnarray}
We have rewritten the effective quantities in terms of the quantities in a vacuum ($V_{\alpha}=0$),
$\sin^{2}(2\theta) = 4 F_{\alpha}^{2} v^{2} / (4 F_{\alpha}^{2} v^{2} + M^{2})$
and  
$\Delta m^{2}/2p = M \sqrt{4 F_{\alpha}^{2} v^{2} + M^{2}}$.

The SM particles contribute to $V_{\alpha}$ through the weak interaction\,\cite{1988NuPhB.307..924N}.
We expand $V_{\alpha}$ in terms of $p/m_{\rm W/Z}, T/m_{\rm W/Z}$ ($m_{\rm W/Z}$ is the mass of W/Z-boson).
The leading contribution is proportional to the lepton number $n_{L_{\alpha}}-n_{\bar L_{\alpha}}$,
\footnote{In general the baryon number also makes a leading contribution to the potential $V_{\alpha}$\,\cite{1988NuPhB.307..924N}. 
The baryon number of the Universe, however, is constrained to be negligibly small $n_{\rm b}/s \simeq 8\times10^{-11}$ ($s$ is entropy density) by the CMB\,\cite{Komatsu2011}.}
\begin{eqnarray}
V^{0}_{\alpha} &=& \pm \frac{1}{\sqrt{2}} G_{\rm F} {\big [}(1+ 4 x_{\rm W}) (n_{l_{\alpha}} - n_{{\bar l}_{\alpha}}) - 
(1 - 4 x_{\rm W}) {\textstyle \sum}_{\beta \neq \alpha}  (n_{l_{\beta}} - n_{{\bar l}_{\beta}}) \notag \\
&& \qquad \quad  + 4  (n_{\nu_{\alpha}} - n_{{\bar \nu}_{\alpha}}) + 2
{\textstyle \sum}_{\beta \neq \alpha} (n_{\nu_{\beta}} - n_{{\bar \nu}_{\beta}}) {\big ]} \,,
\label{eq:potential0th}
\end{eqnarray}
with the Fermi constant $G_{\rm F}$ and the Weinberg angle $x_{\rm W} \equiv \sin^{2}\theta_{\rm W} \simeq 0.23$.
The potential takes an opposite sign for the neutrino ($+$) and anti-neutrino ($-$).
If the lepton number is of the same order as the baryon asymmetry due to, for instance, the sphaleron process\,\cite{1983PhRvD..28.2019M, 1984PhRvD..30.2212K}, $V^{0}_{\alpha}$ is negligibly small.
Therefore we should consider the next leading contribution, which is proportional to the lepton energy density plus pressure,
\begin{eqnarray}
V^{1}_{\alpha} &=& 
-\frac{8 \sqrt{2} G_{\rm F} p}{3 m_{\rm Z}^2} (\rho_{\nu_{\alpha}} + \rho_{{\bar \nu}_{\alpha}})
-\frac{2 \sqrt{2} G_{\rm F} p}{m_{\rm W}^2} (\rho_{l_{\alpha}} + P_{l_{\alpha}} + \rho_{{\bar l}_{\alpha}} + P_{{\bar l}_{\alpha}}) \,.
\label{eq:potential1st}
\end{eqnarray} 
The next leading contribution has the same sign for the neutrino and anti-neutrino.

As we can see from eqs.\,(\ref{eq:effectiveangle}), (\ref{eq:potential0th}), and (\ref{eq:potential1st}), the leading potential $V^{0}$ may cause a resonance in the active-sterile mixing, while the next leading potential $V^{1}$ always suppresses the mixing.
If large lepton asymmetry is produced after the sphaleron process (by, e.g., the degenerate right-handed neutrino oscillation\,\cite{2008JHEP...08..008S} or the Affleck-Dine mechanism\,\cite{2002PhRvD..66d3516K}), it may result in the resonant production of the sterile neutrino, which can account for the observed DM mass density\,\cite{1999PhRvL..82.2832S, 2008JCAP...06..031L}.
In light of the reported X-ray anomaly around 3.5 keV, the resonantly produced sterile neutrino is revisited and its implications for structure formation is investigated in ref.\,\cite{Abazajian2014, 2015arXiv151204544B, 2015arXiv151204548H}.
In the present paper, we assume small lepton asymmetry and hence $V^{0}\simeq0$.

With the next leading potential, the effective mixing angle can be rewritten as,
\begin{eqnarray}
\sin^{2}(2\theta_{\rm eff}) = \frac{\sin^{2}(2\theta)}{\sin^{2}(2\theta)+\left\{ \cos(2\theta)+0.02\zeta(p/T)^{2}(T/100\,{\rm MeV})^{6}(1\,{\rm keV}^{2}/\Delta m^{2}) \right\}^{2}} \,,
\label{eq:tempdepangle}
\end{eqnarray}
where we assume among the charged leptons only the electron is relativistic and makes a relevant contribution at the temperature of interest ($T \sim $100$\,{\rm MeV}$ as we explain below) and $\zeta=1.0$ for $\nu_{\rm e}$ and $\zeta=0.3$ for $\nu_{\mu/\tau}$.
This implies that the sterile neutrino is produced efficiently at temperatures below $100\,{\rm MeV}(p/T)^{-1/3}(\Delta m^{2}/1\,{\rm keV}^{2})^{1/6}$.

The Boltzmann equation of the sterile neutrino can be written as\,\cite{Dodelson1994, 2002APh....16..339D, 2001PhRvD..64b3501A},
\begin{eqnarray}
\frac{\partial}{\partial t} f_{\rm s} - Hp \frac{\partial}{\partial p} f_{\rm s} = \sin^{2}(\theta_{\rm eff}) \Gamma f_{\rm th} \,,
\end{eqnarray}
where $f_{\rm th} = (\exp(p/T)+1)^{-1}$ is the Fermi-Dirac distribution.
The collision rate $\Gamma_\alpha$ is proportional to the lepton energy density plus pressure\,\cite{1992NuPhB.373..498E, 1994PhRvD..49.2710M}.
After summing up all contributions, we obtain
\begin{eqnarray}
\Gamma_\alpha = \frac{7\pi}{540} G_{\rm F}^{2} T^{4} p \times
\begin{cases}
40 x_{\rm W}^2 + 20 x_{\rm W} + 25 & \text{for} ~ \nu_{\rm e} \\
40 x_{\rm W}^2 - 20 x_{\rm W} + 25 & \text{for} ~ \nu_{\mu/\tau}
\end{cases},
\end{eqnarray}
where we again assume only the electron is relativistic and makes a relevant contribution.
The number of collision per Hubble time increases with the temperature as $\Gamma_\alpha/H \propto \sin^{2}(2\theta)(p/T)T^{3}$.
From this and eq.\,(\ref{eq:tempdepangle}), we can find that the sterile neutrino is mainly produced at $T \sim 100\,{\rm MeV}(p/T)^{-1/3}(\Delta m^{2}/1\,{\rm keV}^{2})^{1/6}$.

Now let us consider the momentum distribution of the resultant sterile neutrino.
At sterile neutrino production, the number of production reactions per Hubble time is given by 
$\sin^2 (\theta_{\rm eff}) \Gamma_\alpha/H \propto \sin^{2}(\theta) (\Delta m^{2}/1\,{\rm keV}^{2})^{1/2}$ 
and is independent of the momentum.
Therefore the resultant momentum distribution of the sterile neutrino is proportional to the Fermi-Dirac distribution, $f_{\rm s} \propto \sin^{2}(2\theta) (\Delta m^{2})^{1/2} f_{\rm th}$\,\cite{Dodelson1994, 2001PhRvD..64b3501A}.
\footnote{\label{fn:relicofsterilenu} This estimate suffers from ignorance of the quark-hadron phase transition. 
This may suppress the estimate of the relic abundance at most by a factor of two\,\cite{2002PhRvD..66b3526A, 2006JHEP...06..053A, 2007JHEP...01..091A}. 
In addition, the induced momentum dependence of $f_{\rm s}/f_{\rm th}$ may change the resultant power spectrum by at most $20\%$\,\cite{2006PhRvD..73f3506A}.}
Finally we obtain the relic density of the sterile neutrino,
\begin{eqnarray}
\Omega_{\rm s}h^{2} \simeq 0.07 \left(\frac{\sin^{2}(2\theta)}{10^{-9}}\right) \left( \frac{M}{7\,{\rm keV}} \right)^{2} \,,
\label{eq:relic}
\end{eqnarray}
where we have used $\Delta m^{2} \simeq M^{2}$ for $M \gg F_{\alpha} v$.

The sterile neutrino introduced in the above has an interesting astrophysical implication.
The dominant decay mode of the sterile neutrino is $\nu_{\rm R} \to 3 \nu_{\rm L}/{\bar \nu}_{\rm L}$ with a decay rate of \cite{1984PhRvD..30.2422G}
\begin{eqnarray}
\Gamma_{\nu_{\rm R} \to 3\nu_{\rm L}/{\bar \nu}_{\rm L}} &=& \frac{G_{\rm F}^{2}}{384\pi^{3}} \sin^{2} (2 \theta) \, M^{5} \\
&\simeq& 2.9 \times 10^{-25}\,{\rm s}^{-1} \left( \frac{\sin^{2} (2 \theta)}{10^{-9}} \right) \left( \frac{M}{7\,{\rm keV}} \right)^{5} \,.
\end{eqnarray}
The lifetime of the sterile neutrino is much longer than the age of the Universe ($\sim 10^{17}\,{\rm s}$). 
This ensures that the sterile neutrino produced in the early Universe still exists at present; it is a good candidate of (or a part of) DM.
The radiative decay of the sterile neutrino $\nu_{\rm R} \to \gamma + \nu_{\rm L}/{\bar \nu}_{\rm L}$ is subdominant but has an important implication for the observed X-ray anomaly around $3.5$\,keV. 
The radiative decay rate is given by\,\cite{1982PhRvD..25..766P}
\begin{eqnarray}
\Gamma_{\nu_{\rm R} \to \gamma + \nu_{\rm L}/{\bar \nu}_{\rm L}} &=& \frac{9 G_{\rm F}^{2} \alpha}{1024\pi^{4}} \sin^{2} (2 \theta) \, M^{5} \\
&\simeq& 2.3 \times 10^{-27}\,{\rm s}^{-1} \left( \frac{\sin^{2} (2 \theta)}{10^{-9}} \right) \left( \frac{M}{7\,{\rm keV}} \right)^{5} \,,
\label{eq:decayrate}
\end{eqnarray}
where $\alpha$ is the fine structure constant of quantum electrodynamics. 
The sterile neutrino decay in an overdense region contributes to the observed X-ray flux,
\begin{eqnarray}
F_{\rm s} &=& \frac{1}{4\pi(1+z)^{3}} \frac{\Sigma_{\rm s}}{M} \Omega_{\rm fov} \Gamma_{\nu_{\rm R} \to \gamma + \nu_{\rm L}/{\bar \nu}_{\rm L}} \\
&\simeq& \frac{4 \times 10^{-5}}{(1+z)^{3}} \, \frac{\rm cts}{\rm s\,cm^{2}}
\left( \frac{\Sigma_{\rm DM}}{500\,M_{\odot}/{\rm pc}^{2}} \right) 
\left( \frac{\Omega_{\rm fov}}{500\,{\rm arcmin}^{2}} \right)
\left( \frac{\sin^{2} (2 \theta)}{10^{-9}} \right)^{2}
\left( \frac{M}{7\,{\rm keV}} \right)^{6} \,,
\end{eqnarray}
where $\Sigma_{\rm s/{\rm DM}}$ and $\Omega_{\rm fov}$ are the column density of the sterile neutrino/DM and the field of view of the target object, respectively.
In the second line, we have used $\Sigma_{\rm s}=\Sigma_{\rm DM}\Omega_{\rm s}/\Omega_{\rm DM}$ and eqs.\,(\ref{eq:relic}) and (\ref{eq:decayrate}).
Following the analysis in ref.\,\cite{Boyarsky2014}, we adopt a mixing angle of $\sin^{2}(2\theta) \simeq 3 \text{--} 9 \times 10^{-10}$ to explain the observed X-ray anomaly around $3.5\,{\rm keV}$. 
In this case the sterile neutrino accounts for $20\text{--}60\%$ of the whole DM mass density. 
Here we have used the observed flux from the Andromeda galaxy (on-center) $F_{\rm s}=4.9^{+1.6}_{-1.3} \times 10^{-6}\, 
{\rm cts}/{\rm s}/{\rm cm}^{2}$ and the column density of DM $\Sigma_{\rm DM} \simeq 200 \text{--} 600\,M_{\odot}/{\rm pc}^{2}$.
We have also taken into account the possible suppression in the sterile neutrino relic density $\Omega_{\rm s}$ by a factor of two (see footnote\,\ref{fn:relicofsterilenu}). 
For the Perseus cluster, the observed X-ray flux and the column density are $F_{\rm s}=7.0^{+2.6}_{-2.6} \times 10^{-6}\, {\rm cts}/{\rm s}/{\rm cm}^{2}$  and $\Sigma_{\rm DM} \simeq 100 \text{--} 600\,M_{\odot}/{\rm pc}^{2}$, respectively\,\cite{Boyarsky2014}. 
The observed flux prefers a mixing angle of $\sin^{2}(2\theta) \simeq 0.3 \text{--} 2 \times 10^{-9}$, which results in $r_{\rm warm}=\Omega_{\rm s}/\Omega_{\rm DM}\simeq 0.2 \text{--} 1$. 
Finally, by combining these two analyses, we find that $\sin^{2}(2\theta) \simeq 3 \text{--} 9\times 10^{-10}$ and $r_{\rm warm} \simeq 0.2 \text{--} 0.6$ can explain X-ray flux both from the Andromeda galaxy and from the Perseus galaxy cluster.
However, let us stress that the observations of the Lyman-$\alpha$ forests disfavors $r_{\rm warm}>0.6$ at this mass\,\cite{Boyarsky2009}.

In the following sections, we do not specify what accounts for the rest of the whole DM mass density but only assume that it is composed of some stable and cold particles. Interestingly, if the remainder is thermal relic WIMPs, the annihilation of the WIMPs can naturally produce sufficient $\gamma$-rays to resolve the observed anomaly in the Galactic center\,\cite{2011PhRvD..84l3005H, 2013PhRvD..88h3521G, 2014arXiv1402.4090A, 2014arXiv1403.1987L}. As an example, we consider the case that WIMPs annihilate into $b\bar{b}$. From the figure\,6 in ref.\,\cite{2011PhRvD..84l3005H}, the annihilation cross section (multiplied by the relative velocity) required to explain the $\gamma$-ray anomaly is $\langle \sigma v \rangle \simeq 0.7 \text{--} 1.6 \times 10^{-26}\,{\rm cm^3/s}$ (depending on the mass of WIMP, here taken to be $m_{\rm WIMP} \simeq 20 \text{--} 40\,{\rm GeV}$) if all of the DM is composed of WIMPs. However, since sterile neutrinos comprise $25\%$ or $50\%$ of the DM abundance in our models, the annihilation cross section must increase to some $\langle \sigma v \rangle \simeq 4\text{--}9\,{\rm cm^3/s}$ in order to explain the observed $\gamma$-ray flux (provided the central DM density remains the same). For thermally decoupling WIMPs, the increased cross section also results in a reduced relic abundance. Numerically, the WIMPs contribute approximately $0.3\text{--}0.6$ towards the DM abundance, naturally bringing the WIMP + sterile neutrino abundance to $\Omega_{\rm WIMP} + \Omega_{\rm s} \sim \Omega_{\rm DM}$.

In the above estimate, we focus on the case that WIMPs annihilate into $b\bar{b}$. 
One may wonder if the result depends on the assumed DM mass density profile and/or the annihilation channel of WIMPs. 
Let us discuss this point further. 
In ref.\,\cite{2011PhRvD..84l3005H}, the inner slope of the DM mass density profile is assumed to be $\gamma=1.3$. 
If one takes $\gamma=1.2$, the required annihilation cross section becomes roughly three times larger when all of DM particles are composed of WIMPs\,\cite{2013PhRvD..88h3521G}. 
Following the discussion above, we find that WIMPs account for at least 80\% of the whole DM mass density.
As long as we consider only prompt photons from the WIMP annihilation, the required annihilation cross section is almost insensitive to the annihilation channel\,\cite{2011PhRvD..84l3005H}. 
The annihilation channels becomes relevant when we incorporate secondary photons, for instance, photons from bremsstrahlung of prompt electrons\,\cite{2014arXiv1402.4090A, 2014arXiv1403.1987L}. 
The required annihilation cross section does not change drastically for the $b\bar{b}$ channel, while it becomes roughly three times smaller for the lepton ($1/3$ $e^{+}e^{-}$, $1/3$ $\mu^{+}\mu^{-}$, $1/3$ $\tau^{+}\tau^{-}$) channel. 
From these discussions, we expect that the preferred fraction $\Omega_{\rm WIMP}/\Omega_{\rm DM}$ is uncertain roughly by a factor of three.

\section{Linear and non-linear evolution}
\label{sec:evolution}
\subsection{Linear growth}
\label{sec:linear}
We consider models in which DM consists of the NRP sterile neutrino and some cold and stable particles.
Free-streaming of the NRP sterile neutrino suppresses growth of the matter density fluctuation. 
As discussed in section\,\ref{sec:neutrino}, the ratio of the sterile neutrino (warm component) to the whole DM mass density
is $r_{\rm warm} \simeq 0.2\text{--}0.6$ when the mixing angle is $\sin^{2}(2\theta) \simeq 3 \text{--} 9\times 10^{-10}$.
Then the radiative decay of the sterile neutrino can explain the X-ray anomaly around $3.5$\,keV reported in the recent XMM Newton and Chandra observations. 
We focus on the two models with $r_{\rm warm} = 0.25$ and $r_{\rm warm} = 0.50$, respectively.

For comparison, we calculate the evolution of the matter density fluctuations in the CDM model and a WDM model with the thermal relic mass $m_{\rm WDM} = 2.4\,{\rm keV}$ in addition to the MDM models discussed above. 
We use the public code \verb|CAMB|\,\cite{Lewis2000} with suitable modification, through which NRP sterile neutrino is incorporated. 
Details of the modification are as follows. 
In order to take the effect of free-streaming of DM particles into account, we use the method developed for the massive neutrino\,\cite{Ma1995, Lewis2002}. 
In the WDM model, the WDM particles follow the Fermi-Dirac distribution. 
In the MDM models, the Fermi-Dirac distribution of the massive neutrino is replaced by the momentum distribution of the NRP sterile neutrino (see the previous section). 
We adopt cosmological parameters of WMAP+BAO+$H_0$ Mean in ref.\,\cite{Komatsu2011}: $100\Omega_{\rm b} h^2 = 2.255$, $\Omega_{\rm CDM} h^2 = 0.1126$, $\Omega_\Lambda = 0.725$, $n_{\rm s} = 0.968$, $\tau = 0.088$, $\Delta_\mathcal{R}^2(k_0) = 2.430 \times 10^{-9}$, $h=0.70$. We set the energy density of the CDM + WDM to be $0.1126$.

\begin{figure}[tbp]
\centering
\includegraphics[width=\hsize]{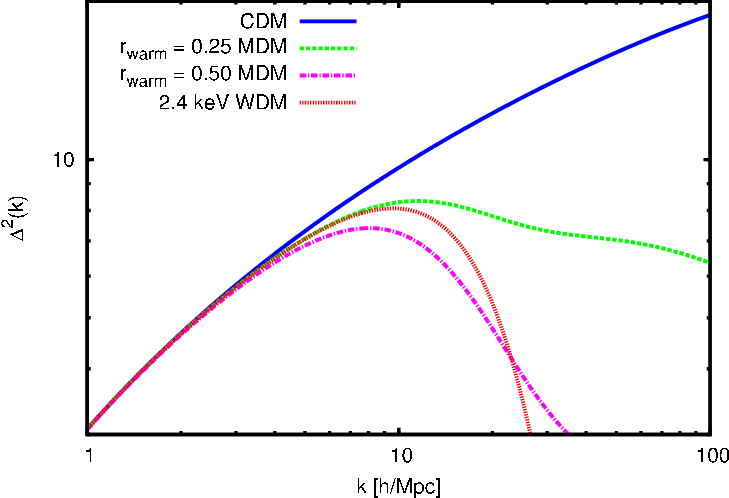}
\hfill
\caption{\label{fig:spectrum} Linear power spectra in the CDM model (blue solid), the MDM models with $r_{\rm warm} = 0.25$ (green dashed) and $r_{\rm warm} = 0.50$ (magenta dash-dotted), and the WDM model with $m_{\rm WDM} = 2.4\,{\rm keV}$ (red dotted). 
Cutoff scales (eq.\,(\ref{eq:jeans})) of the MDM model with $r_{\rm warm} = 0.25$ and the WDM model are identical. 
In the WDM model, strong suppression is seen at wavenumber $k$ larger than $\simeq 10 \,h/{\rm Mpc}$. 
While the MDM models also show suppressions, they are weaker than that in the WDM model. 
This is because the warm component contributes to $25\%$ and $50\%$ of the whole DM mass density in the MDM models.}
\end{figure}

In figure\,\ref{fig:spectrum}, we compare the linear matter power spectra in our selected models.
The MDM models with $r_{\rm warm} = 0.25$ (green dashed line) and $0.50$ (magenta dash-dotted line) result in slightly suppressed linear matter power spectra when compared with that in the CDM model (blue solid line). 
This is due to the free-streaming of the NRP sterile neutrino. 
The MDM models show smaller suppression than the WDM model (red dotted line) because the NRP sterile neutrino contributes to only $25\%$ and $50\%$ of the whole DM mass density.

One may regard the thermal relic WDM model with mass $m_{\rm WDM} = 2.4\,{\rm keV}$ as being a counterpart of the MDM model with $r_{\rm warm} = 0.25$, because the cutoff scales of the linear density fluctuations are identical
in the two models. 
Here, the cutoff scale is defined by the comoving Jeans scale at matter radiation equality $t_{\rm eq}$,
\beqa
k_{\rm J} = a\sqrt{\frac{4 \pi G \rho_0}{\sigma^2}} \Bigg|_{t=t_{\rm eq}} &=& 64 \, {\rm Mpc}^{-1} \left(\frac{m_{\rm WDM}}{2.4\,{\rm keV}}\right)^{4/3} \\ &=& 64 \, {\rm Mpc}^{-1} \left(\frac{M}{7\,{\rm keV}}\right)\left(\frac{0.25}{r_{\rm warm}}\right)^{1/2} \,, \label{eq:jeans}
\eeqa
where $a$ is the scale factor, $G$ is the gravitational constant, $\rho_0$ is the mean matter density, $\sigma^2 \equiv \int_0^\infty v^2 g(v) dv$ is the velocity dispersion of DM particles, and $g(v)$ is the isotropic mass-weighted velocity distribution function normalized as $\int_0^\infty g(v) dv = 1$. 
In the MDM models, $g(v)$ is given by $g(v) = r_{\rm warm} g_{\rm warm}(v) + (1-r_{\rm warm})g_{\rm cold}(v)$, where $g_{\rm cold/warm}(v)$ is the isotropic velocity distribution function of cold/warm component normalized as $\int_0^\infty g_{\rm cold/warm}(v)dv = 1$.

The observations of the Lyman-$\alpha$ forests severely constrain the WDM model but not the MDM model, even though the cutoff scales are identical. 
The recent constraint on the WDM mass derived by ref.\,\cite{Viel2013} is $m_{\rm WDM} > 3.3\,{\rm keV}$ and hence the WDM model in figure\,\ref{fig:spectrum} is indeed excluded. 
However, the constraint is not applicable to the MDM models. 
Reference\,\cite{Boyarsky2009} shows the parameter region that is consistent with the observations of the Lyman-$\alpha$ forests in the MDM models. 
From figure 3 in ref.\,\cite{Boyarsky2009}, we find that our MDM models evade the constraint from the observations of the Lyman-$\alpha$ forests.

The characteristic cutoff length scale given by eq.\,(\ref{eq:jeans}) implies a possibility that the MDM models resolve the small-scale crisis. 
The Jeans mass corresponding to $k_{\rm J}$ is 
\beq
M_{\rm J} = \frac{4\pi}{3}\left(\frac{2\pi}{k_{\rm J}}\right)^3\rho_0 \simeq 8.7 \times 10^7 \left(\frac{64 {\rm Mpc}^{-1}}{k_{\rm J}}\right)^3 \, M_\odot/h \,. \label{eq:jeansmass}
\eeq
Formation of halos with masses below $M_{\rm J}$ is expected to be suppressed. 
Since $M_{\rm J}$ is 
at the subgalactic scale,  
the so-called small-scale crisis may be alleviated in the MDM models. 
This motivates us to perform simulations to investigate how strongly the formation of subhalos is suppressed in Galactic-size halos.

\subsection{Numerical simulations}
\label{sec:mf}
We use the parallel Tree-Particle Mesh code \verb|GADGET-2|\,\cite{Springel2005}. 
We use the linear matter power spectra shown in figure\,\ref{fig:spectrum}. 
The simulated volume is $L^3=(5\,{\rm Mpc}/h)^3$. 
We employ $N = 512^3$ particles with uniform mass of $m_{\rm part} \simeq 7.1 \times 10^4\,M_\odot/h$ and set the gravitational softening length $\epsilon = 500\,{\rm pc}/h$. 
The initial redshift of our simulation is $z = 14$.
In order to compensate for the relatively small box size, we generate six realizations for each model. 
Samples of the initial fluctuations are identical among the DM models in each realization 
except for the linear matter power spectra.
On the other hand, they are different among the six realizations.
One may wonder if the relatively small box size may affect internal properties of simulated halos significantly.
We address this point in appendix\,\ref{sec:resolution}.
We also discuss the effect of the gravitational softening length in the appendix.

\begin{figure}[tbp]
\centering
\includegraphics[width=\hsize]{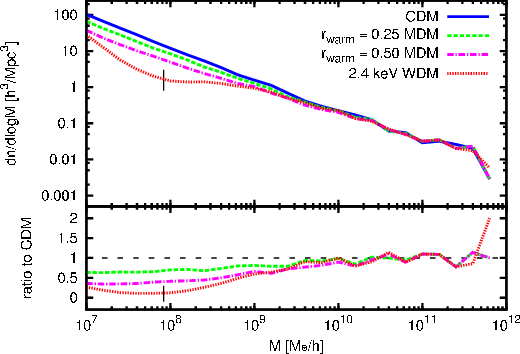}
\hfill
\caption{\label{fig:mf} Upper panel: Halo mass functions in the CDM model, the MDM models, and the thermal WDM model, as labeled. 
The mass function in each model is averaged over six realizations.
The line colors and types are the same as in figure\,\ref{fig:spectrum}. 
The black vertical line represents the mass $M_{\rm lim}$, below which the mass function in the WDM model is numerically compromised because of the discreteness effect. 
Lower panel: Ratios of the mass functions in the MDM models with $r_{\rm warm} = 0.25$ (green dashed) and $0.50$ (magenta dash-dotted) and the WDM model (red dotted) to that in the CDM model. 
The black vertical line also shows $M_{\rm lim}$. 
Below $\sim 10^{10}\,M_\odot/h$, the abundance of small halos is suppressed in the MDM models and the WDM model 
compared to the CDM model. 
The suppression is stronger in the WDM model than in the MDM models.}
\end{figure}

We show the halo mass functions from the DM only $N$-body simulations 
in figure\,\ref{fig:mf}. 
Halos are identified by Friends-Of-Friends (FOF) method\,\cite{Davis1985}. 
The linking length is $0.2$ in units of the mean interparticle distance. 
Hereafter, unless explicitly denoted, halo mass is the FOF mass, which roughly coincides with the virial mass $M_{200}$.
In figure\,\ref{fig:mf}, the number of halos with masses below $\sim 10^{10}\,M_\odot/h$ is suppressed in the MDM models (green dashed for $r_{\rm warm} = 0.25$ and magenta dash-dotted for $0.50$) compared to that in the CDM model (blue solid). 
In the WDM model (red dotted), even stronger suppression is seen below the same mass.
Note that a numerical discreteness effect is seen in the WDM model. 
In refs.\,\cite{Wang2007, Kamada2013}, it is reported that halos with masses below
\beq
M_{\rm lim} = 10.1 \times \rho_0 \bar{d} k_{\rm peak}^{-2} = 8.3 \times 10^7 M_\odot/h 
\left(\frac{L}{5\,{\rm Mpc}/h} \right) 
\left(\frac{512}{N^{1/3}} \right) 
\left( \frac{9.5\,h/{\rm Mpc}}{k_{\rm peak}} \right)^2 \label{eq:mlim} %8.3097e7 
\eeq
may be seeded by discrete particle effects and hence the mass function below this ``critical'' mass is unreliable in hot/warm DM model. 
Here, 
$\bar{d} = L/N^{1/3}$ is the mean interparticle distance 
and 
$k_{\rm peak}$ is the wavenumber at the maximum of $\Delta(k)$ (see figure\,\ref{fig:spectrum}). 
The black vertical lines in figure\,\ref{fig:mf} indicate $M_{\rm lim}$ in the WDM model. 
Thus, the upturn seen to the left of the black line in the WDM model is likely unphysical. 
While it is unclear how the discreteness effect changes the mass function in MDM models, it seems to be insignificant because no clear upturn is found in the lower panel of figure\,\ref{fig:mf}.

\subsection{Subgalactic scale}
\label{sec:cvf}
\begin{figure}[tbp]
\centering
\includegraphics[width = 0.55\hsize]{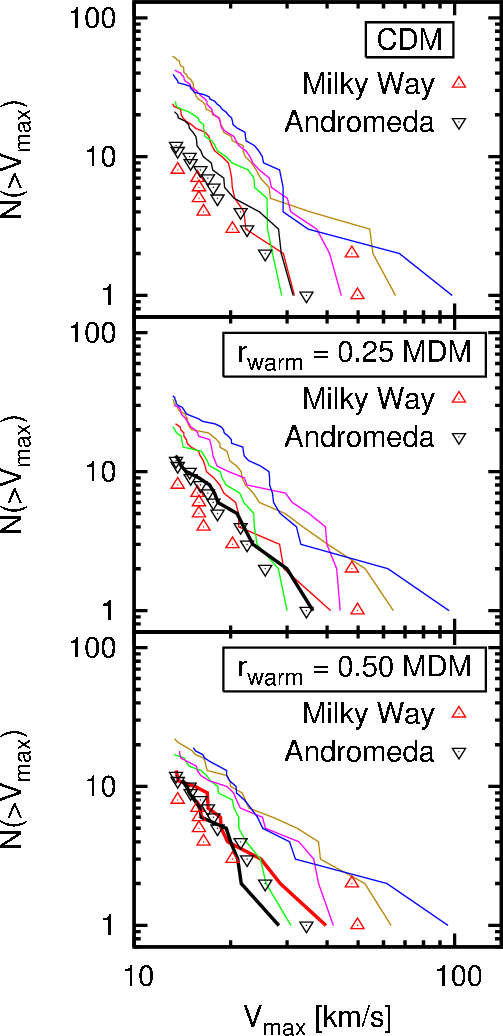}
\caption{\label{fig:cvf} CVFs measured from observation and our simulations. 
The up-triangles are for the Milky Way satellites, and the down-triangles represent the Andromeda satellites. 
The solid lines are the CVFs in our simulations. 
The different line colors correspond to different halo masses: $\sim 2.6\times 10^{12} M_\odot$ (magenta), $\sim 2.1\times 10^{12} M_\odot$ (blue), $\sim 1.8\times 10^{12} M_\odot$ (dark-yellow), $\sim 1.7\times 10^{12} M_\odot$ (green), $\sim 1.5\times 10^{12} M_\odot$ (red),  and $\sim 1.4\times 10^{12} M_\odot$ (black).
The top panel is a comparison between the CVFs in the CDM model and those observed. 
There are too many simulated subhalos than the observed satellites at small $V_{\rm max}$. 
The other panels show comparisons between the CVFs in the MDM models (middle for $r_{\rm warm} = 0.25$, and bottom for $0.50$) and those observed. 
The bold lines show halos whose CVFs are close to those observed.}
\end{figure}

The subgalactic scale Jeans mass (eq.\,(\ref{eq:jeansmass})) and the smaller number of low mass halos (figure\,\ref{fig:mf}) imply that the MDM models have smaller numbers of subhalos and hence possibly resolve the missing satellite problem. 
We compare the cumulative maximum circular velocity functions (CVFs) of our simulated halos with those observed in figure\,\ref{fig:cvf}.
We use \verb|SUBFIND|\,\cite{Springel2001} to locate subhalos.
The maximum circular velocity is defined by
\beq
V_{\rm max} = \max_{r} \sqrt{\frac{GM(<r)}{r}} \,,  \label{eq:vmax}
\eeq
where $M(<r)$ is the enclosed mass within the radius $r$ from the subhalo center. 
The direct observable is the line-of-sight stellar velocity rather than the maximum circular velocity. 
Thus, we assume isotropic velocity distribution and adopt conventional simple conversion\,\cite{Moore1999}
\beq
V_{\rm max} = \sqrt{3}\sigma_{\rm los} \,, \label{eq:los}
\eeq
where $\sigma_{\rm los}$ is the line-of-sight stellar velocity dispersion.

The conversion factor $\sqrt{3}$ (e.g., eq.\,(1) in ref.\,\cite{Wolf2010}) yields a lower bound because it gives a fair estimate of the circular velocity at the half-light radius (not the radius $R_{\rm max}$ at which the circular velocity reaches its maximum). 
The error in eq.\,(\ref{eq:los}) is typically less than $\sim10\text{--}20\%$\,\cite{Klypin1999}, and thus the choice of either eq.\,(\ref{eq:vmax}) or eq.\,(\ref{eq:los}) does not affect the results significantly.
On the other hand, the conversion factor is estimated at $\simeq 2\text{--}3$ by assuming a relation between $R_{\rm max}$ and $V_{\rm max}$ found for galaxy cluster size halos\,\cite{Penarrubia2008}. 
We discuss the implication of the larger conversion factor in section\,\ref{sec:conclusion}. 
Ideally, $\sigma_{\rm los}$ should be measured directly from the simulation data. 
However, $\sigma_{\rm los}$ is observationally measured in $\sim 1\,{\rm kpc}$ regions from the center of satellite galaxies. Since our simulations do not resolve such an inner region, we can not compute $\sigma_{\rm los}$ directly in the same manner as the observations. 

Observational data (triangles in figure\,\ref{fig:cvf}) are taken mainly from ref.\,\cite{McConnachie2012}. 
We use satellite galaxies associated with the Milky Way (MW) or the Andromeda galaxy in ref.\,\cite{McConnachie2012}. 
We regard the maxima of the rotation curve reported in ref.\,\cite{McConnachie2012} as the maximum circular velocities of LMC, NGC 205, 147, and 185 (see appendix\,\ref{sec:velconv}). 
For IC 10 associated with the Andromeda galaxy, we use the peak of $\rm H_I$ disk rotation curve in ref.\,\cite{Wilcots1998} 
as $V_{\rm max}$. 
We apply the approximation (eq.\,(\ref{eq:los})) to the other satellites. 
The line-of-sight stellar velocity dispersion of the satellites associated with the MW is taken from ref.\,\cite{McConnachie2012}. 
For the satellites associated with the Andromeda galaxy, we use the line-of-sight stellar velocity of refs.\,\cite{Ho2012} (And II), \cite{Collins2014} (And XVI, XXI), \cite{McConnachie2012} (And XII), and those in ref.\,\cite{Collins2013} (for the other satellites). 
Pisces II and And XXIV are not included in figure\,\ref{fig:cvf} because their velocity dispersions are not known. 
Thus, the observed CVFs will change slightly if the two satellites are included with updated velocity measurements. 

From simulations, we extract halos with masses between $7.0\times 10^{11}\,M_\odot$ and $2.7 \times 10^{12}\,M_\odot$. We classify halos whose masses are larger than $7.1 \times 10^{11}\,M_\odot$\,\cite{Klypin2002} 
\footnote{The lowest mass estimated in their work. 
In order to model the MW halo, they use Navarro-Frenk-White (NFW) profile\,\cite{Navarro1996}, which is established in the CDM model. 
The halo density profiles in the CDM model and the MDM models are similar as seen in appendix \ref{sec:densityprofile}. Thus, their estimate is also valid for the MDM model halos.} 
and smaller than $2.7 \times 10^{12}\,M_\odot$\,\cite{Watkins2010} as candidates of the MW, and halos whose masses are larger than $7.0 \times 10^{11}\,M_\odot$\,\cite{Evans2000} and smaller than $2.1 \times 10^{12}\,M_\odot$\,\cite{Fardal2013} as candidates of the Andromeda galaxy. 
Within these host halos, we identify subhalos by \verb|SUBFIND| algorithm\,\cite{Springel2001} and calculate $V_{\rm max}$ from eq.\,(\ref{eq:vmax}) where the density peak is regarded as the center of subhalo. 
We assume that the most massive subhalo, or ``background halo'', corresponds to the MW or the Andromeda galaxy itself. 
Thus, figure\,\ref{fig:cvf} does not include such a background halo. 

We do not use all the satellite galaxies or the subhalos explained above because of several limitations.
First, we only use satellite galaxies and subhalos whose $V_{\rm max}$'s are larger than $13.16\,{\rm km/s}$. 
More than half of the satellites were not discovered until Sloan Digital Sky Survey (SDSS).
Since SDSS does not cover the whole sky, more satellite galaxies will be discovered if the whole sky is surveyed. 
However, there are no MW satellites discovered by SDSS whose $V_{\rm max}$'s are larger than $13.16\,{\rm km/s}$. This implies that using satellite galaxies within this range allows robust comparison between observational data and predictions of our simulations. 
The Andromeda satellites in this range are also expected to be large enough to be discovered completely. 
In addition, simulated subhalos in this velocity range are well resolved (see also appendix\,\ref{sec:resolution}). 
All such subhalos have masses larger than $M_{\rm lim}$ (eq.\,(\ref{eq:mlim})), large enough $R_{\rm max} > 5\times\epsilon$ except for a few subhalos (at most three in each halo in the CDM model, and at most two in the MDM models), and at least 985 particles within $R_{\rm max}$. 
Hereafter, we only consider satellites and subhalos with $V_{\rm max}>13.16\,{\rm km/s}$. 

Second, we only use satellite galaxies and subhalos whose distances from the host halo centers are smaller than $269\,{\rm kpc}$ and larger than $50\,{\rm kpc}$. 
The furthest observed satellite galaxy, And VI, is $269\,{\rm kpc}$ distant from the Andromeda center. 
Subhalos within $50\,{\rm kpc}$ are possibly affected by the gravitational softening of our simulations\,\cite{2005MNRAS.359.1537R}. 
In the regions close to the host halo center, the subhalos are vulnerable to tidal disruption\,\cite{2010MNRAS.406.1290P, 2014ApJ...786...87B, 2014MNRAS.438.1466A}. 
This effect may be enhanced artificially by the finite gravitational softening length.
We confirmed that the tidal force from the host halo is weaker than (possibly softened) gravitaional force to subhalo center at $R_{\rm max}$ for subhalos beyond $50\,{\rm kpc}$ from the host halo centers.
We consider satellite galaxies and subhalos that satisfy the above criteria as those in the classical dwarf range.

The top panel of figure\,\ref{fig:cvf} clearly shows that there are at least a factor of two difference between the simulated CVFs and those observed at small $V_{\rm max}$ in the CDM model. 
On the other hand, in the MDM models with $r_{\rm warm} = 0.25$ (middle panel) and $r_{\rm warm} = 0.50$ (bottom panel), the overall shapes of the black line in the middle panel and the red and the black lines in the bottom panel agree with the observed CVFs.
In figure\,\ref{fig:cvf}, the halo-to-halo scatter of the CVF is shown in each model as reported in refs.\,\cite{Springel2008, 2009ApJ...696.2115I, 2010MNRAS.406..896B}. 
One may wonder if this implies that even in the CDM model, we have a chance to find a halo whose CVF is consistent with the observational data when we generate more realizations.
As we discuss in detail in appendix\,\ref{sec:cumulative}, the cumulative number of subhalos in the classical dwarf range appears to follow a Negative Binomial distribution with mean $\langle N \rangle \simeq 34$ and standard deviation $\sigma \simeq 8.5$ in the CDM model as suggested in ref.\,\cite{2010MNRAS.406..896B}. 
In this case, in only one out of ten thousand halos, the subhalo number in the classical dwarf range is equal to or less than observed.
On the other hand, within the six simulated halos in the MDM models, there are several halos whose CVFs are consistent with the observational data. 
Therefore, the MDM models can mitigate the small-scale crisis siginificatly.

\begin{figure}[tbp]
\centering
\includegraphics[width = 0.6\hsize]{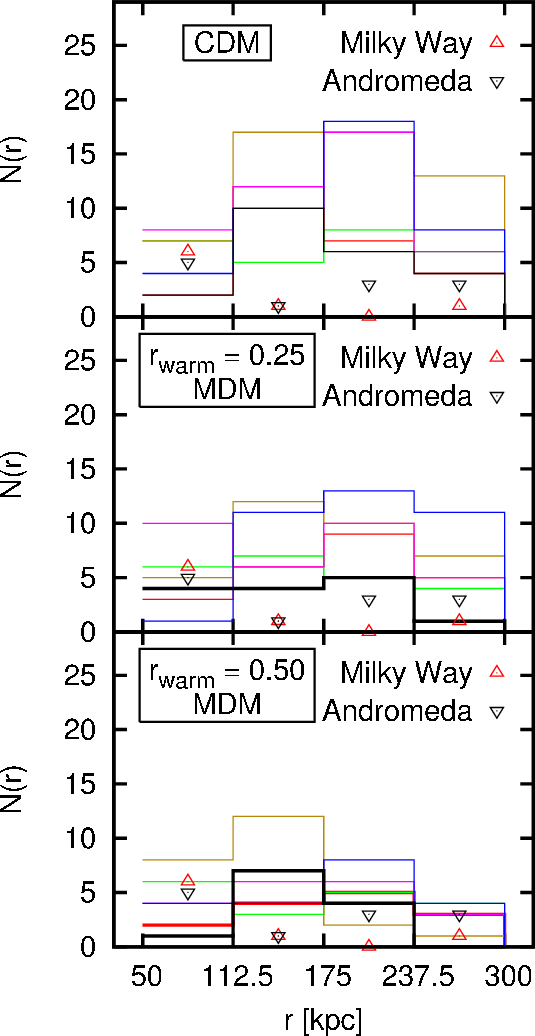}
\caption{\label{fig:radis} Radial distributions of satellite galaxies in the CDM model (top panel) and the MDM models with $r_{\rm warm} = 0.25$ (middle panel) and $0.50$ (bottom panel).
We divide the distance from the halo center between $50\,{\rm kpc}$ and $300\,{\rm kpc}$ into 4 bins. 
The up-triangles are for the MW satellites, and the down-triangles represent the Andromeda satellites. 
The lines show the radial distributions of subhalos in our simulated halos. 
The line colors are the same as those in figure\,\ref{fig:cvf}: $\sim 2.6\times 10^{12} M_\odot$ (magenta), $\sim 2.1\times 10^{12} M_\odot$ (blue), $\sim 1.8\times 10^{12} M_\odot$ (dark-yellow), $\sim 1.7\times 10^{12} M_\odot$ (green), $\sim 1.5\times 10^{12} M_\odot$ (red),  and $\sim 1.4\times 10^{12} M_\odot$ (black).
The bold lines indicate the halos whose CVFs are consistent with those observed.}
\end{figure}

The mitigation of the small-scale crisis can also be found in the radial distributions of satellites shown in figure\,\ref{fig:radis}.
In the plot, the same satellite galaxies and subhalos as in figure\,\ref{fig:cvf} are used. 
All galactocentric distances of satellite galaxies are taken from ref.\,\cite{McConnachie2012}. 
Note that the distance of the farthest observed satellite galaxy is $269\,{\rm kpc}$. 
The line colors and symbols are set consistent with the halos shown in figure\,\ref{fig:cvf}. 
The bold lines correspond to the halos whose CVFs are consistent with those observed (represented by the black line in the middle panel and the red and the black line in the bottom panel of figure\,\ref{fig:cvf}).

The halo-to-halo scatter is seen in figure\,\ref{fig:radis} again. 
Although there are no halos that have an identical radial distributions of subhalos as those observed, the distributions in the MDM models are more concordant with those observed than the CDM model. 
The numbers of subhalos are closer to those observed in all the radial bins except in $50\text{--}112.5\,{\rm kpc}$. 
Table\,\ref{tab:rss} shows the residual sums of squares (RSSs), $\sum_i (N_{i, \,\rm sim.} - N_{i, \,\rm obs.})^2$, where $N_{i, \,{\rm data}}$ is the number of satellites in the $i$-th bin. 
\begin{table}[tbp]
\centering
\begin{tabular}{|c||c|c|c|c|c|c|}
\hline
host halo mass [$10^{12} M_\odot$]& $\sim 1.5$ & $\sim 1.7$ & $\sim 1.8$ & $\sim 2.6$ & $\sim 2.1$ & $\sim 1.4$ \\
\hline
\hline
CDM v.s. MW & 195 & 106 & 690 & 439 & 458 & 142 \\
\hline
CDM v.s. Andromeda & 147 & 54 & 556 & 335 & 332 & 100 \\
\hline
\hline
$r_{\rm warm} = 0.25$ & & & & & &  \\
MDM v.s. MW & 131 & 70 & 258 & 157 & 394 & {\bf 38} \\
\hline
$r_{\rm warm} = 0.25$ & & & & & &  \\
MDM v.s. Andromeda & 69 & 42 & 186 & 103 & 280 & {\bf 18} \\
\hline
\hline
$r_{\rm warm} = 0.50$ & & & & & &  \\
MDM v.s. MW & {\bf 54} & 38 & 129 & 69 & 86 & {\bf 78} \\
\hline
$r_{\rm warm} = 0.50$ & & & & & &  \\
MDM v.s. Andromeda & {\bf 22} & 10 & 135 & 35 & 36 & {\bf 62} \\
\hline
\end{tabular}
\caption{\label{tab:rss} Comparison of residual sums of squares (RSSs). 
The first row represents the halo masses. 
The other rows show RSSs of the simulated radial distributions of subhalos (rows 2 and 3 in the CDM model, rows 4 and 5 in the MDM model with $r_{\rm warm} = 0.25$, and rows 6 and 7 in the MDM model with $r_{\rm warm} = 0.50$) and those observed (rows 2, 4, and 6 for the MW, and rows 3, 5, and 7 for the Andromeda galaxy). 
The bold numbers indicate halos that are represented by the bold lines in figures\,\ref{fig:cvf} and\,\ref{fig:radis}.}
\end{table}
The RSS is systematically smaller in the MDM models than in the CDM model. 

\section{Discussion and conclusion}
\label{sec:conclusion}
We have studied structure formation in MDM models that can explain the unidentified X-ray line around $3.5\,{\rm keV}$ observed by Chandra and XMM-Newton. 
In MDM models, the NRP sterile neutrino with a mass $M \simeq 7\,{\rm keV}$ composes $\sim 25\%$ and $\sim 50\%$ of the whole DM mass density, and some cold stable particle accounts for the rest. 
The mixing angles of the sterile neutrino are $\sin^2(2\theta) \sim 4.0 \times 10^{-10}$ and $8.0 \times 10^{-10}$, respectively. 
These parameters make the MDM models viable and consistent with the observations of the Lyman-$\alpha$ forest. 

We have shown that several MW-like and Andromeda-like halos (in terms of the halo mass) in the MDM models are consistent with the observed Local Group in terms of the CVF. 
The radial distributions of subhalos in the MDM models are also closer to those observed than in the CDM model. 
Therefore, we conclude that the MDM models explain the reported X-ray line and the observations of the Lyman-$\alpha$ forests and mitigate the missing satellite problem.
Besides, the MDM models possibly explain the $\gamma$-ray anomaly in the Galactic center as discussed in section\,\ref{sec:neutrino}.

Adopting a larger conversion factor of eq.\,(\ref{eq:los}) affects the CVFs of the MW and the Andromeda satellites. 
It shifts the CVFs of the observed satellites horizontally to the right in figure\,\ref{fig:cvf}. 
Since the conversion factor is not necessarily constant for all the satellites, it is non-trivial to judge if the MDM models 
explain the observation. 
For instance, if all the satellites have the same conversion factor of 3 rather than $\sqrt{3}$ (3 is the maximum value reported in ref.\,\cite{Penarrubia2008}), the CVFs of the observed satellites shift to the right by a factor of $\sqrt{3}$ in figure\,\ref{fig:cvf}. 
In this case, even the CDM model can reproduce the observation.

It is important to note that there is halo-to-halo scatter in figures\,\ref{fig:cvf} and\,\ref{fig:radis}. 
The scatter is also discussed in refs.\,\cite{Springel2008, 2009ApJ...696.2115I, 2010MNRAS.406..896B, 2014MNRAS.439...73Y}. 
In ref.\,\cite{2010MNRAS.406..896B}, it is suggested that the subhalo mass function and the CVF follow a Negative Binomial distribution with variance larger than expected from the Poisson error. 
The halo-to-halo scatter in figure\,\ref{fig:cvf} is close to that of the suggested distribution, and thus our results are compatible with ref.\,\cite{2010MNRAS.406..896B} (see discussion in appendix\,\ref{sec:cumulative}). 
The scatter of the CVF and the radial distribution of subhalos can be used to calculate the probability that one of the numerous halos has a subhalo population consistent with the observed satellites.
The probability, however, is $\sim 10^{-4}$ for the CDM and much lareger in the MDM models. 
Thus, by introducing the MDM model, distributions of the CVF and the radial distributions of subhalos provide larger possibilities that a halo has its subhalos consistent with observations.
It may be interesting in the future to investigate and reveal those distributions in the MDM models with larger halo samples.

In order to constrain the MDM models further, additional observational data are required. 
The current constraints from the observations of the Lyman-$\alpha$ forests allow these models since the linear matter power spectra in these models show mild suppressions at subgalactic scales. 
Further studies with future observations of the Lyman-$\alpha$ forests are warranted to test the validity of the MDM models.
Another promising candidate is the $21\,{\rm cm}$ line observation. 
By observing the $21\,{\rm cm}$ emission or absorption due to neutral hydrogen clumps associated with dark matter (sub)halo, one can construct the $\rm H_I$ disk velocity width function\,\cite{Zavala2009, Papastergis2011, Schneider2014, Klypin2014}. 
Also, red-shifted $21\,{\rm cm}$ observations hold promise for probing the matter distribution in the cosmic Dark Ages\,\cite{Sekiguchi2014} where the density fluctuations are still in the linear regime even on small length scales. 
Future observations can hopefully constrain the cosmological models such as the MDM models.

We have used dissipation-less DM $N$-body simulations only, and thus we are not able to examine if the (sub)halos host luminous galaxies. 
The abundance matching technique suggests that the galaxy luminosity is tightly correlated with $V_{\rm max}$\,\cite{Conroy2006}. 
On the other hand, the \emph{too-big-to-fail problem}\,\cite{Boylan2011, Boylan2012} suggests that such a simple connection cannot be applied to satellite galaxies\,\cite{Boylan2012}. 
Ultimately, in order to determine if a (sub)halo hosts a luminous object, simulations including baryonic physics will be needed. 
Baryonic effects such as supernovae feedback and photoevaporation by ultra-violet background radiation influence dynamical and star formation properties of subhalos (e.g., ref.\,\cite{Boylan2012, 2002MNRAS.333..177B, 2009MNRAS.399L.174O}). 
The rotation curve in central region can change significantly by such baryonic processes\,\cite{Governato2014}.

The nature of dark matter particles can be studied also from a particle-physics point of view. 
We interpret the X-ray 3.5 keV line as the decay signal of the sterile neutrino that is non-resonantly produced in the early Universe. 
This interpretation actually suggests a larger mixing angle than in the case with resonantly produced sterile neutrino\,\cite{Abazajian2014, 2015arXiv151204544B, 2015arXiv151204548H}. 
Thus we can test our model, in principle, by measuring the mixing angle in some other way. 
One of the tests is pulsar kicks\,\cite{Kusenko2008}. 
It is observed that some pulsars have large bulk velocities directed along their spin. 
Sterile neutrino can cause this pulsar kick, because anisotropically produced active neutrinos in the presence of a magnetic field are converted to sterile neutrinos that escape from the pulsar without washing out the anisotropy.
The recoil of the sterile neutrinos kicks the pulsar. 
When the sterile neutrino has a few keV mass, the onset of kick is considerably delayed because it takes a long time for large lepton asymmetry, which prevents the active-sterile conversion, to relax to zero through mixing\,\cite{Kusenko2008, 2001PhRvD..64b3501A}. 
The delay time depends on the mass of sterile neutrino and the mixing angle, and thus offers a way to place constraints on these basic quantities (see figure 1 in ref.\,\cite{Kusenko2008}).

\acknowledgments
A.H. thanks the Yukawa Institute for Theoretical Physics at Kyoto University, where this work was initiated during the YITP-W-13-07 on ``Summer School on Astronomy \& Astrophysics 2013''. 
A.H. is supported by Advanced Leading Graduate Course for Photon Science (ALPS) in the University of Tokyo. 
A.K. thanks Kevork Abazajian,  Alexander Kusenko, and Hai-bo Yu for useful discussion. 
We also thank Naoki Yoshida, Shunsaku Horiuchi, and the referees for reading manuscript carefully and giving fruitful suggestions. 
Numerical computations were carried out on Cray XC30 at Center for Computational Astrophysics, National Astronomical Observatory of Japan and on SR16000 at YITP in Kyoto University.

\appendix
\section{Effects of the box size and the gravitational softening length}
\label{sec:resolution}
In this appendix, we examine effects of the box size and the gravitational softening length on internal properties of simulated halos. 
The most significant effect of the box size is expected to be seen in the angular momenta of simulated halos\,\cite{2006MNRAS.370..691P}.
The angular momenta, in turn, may affect the density profiles of halos.
We study the angular momenta of our candidate halos of the MW and the Andromeda galaxy in subsection\,\ref{sec:angularmomenta}.
Subsequently, we show their density profiles in subsection\,\ref{sec:densityprofile}.
There, we also discuss effects of the gravitational softening length.
The gravitational softening length should be sufficiently small such that the $V_{\rm max}$'s of subhalos are well determined. 
Subhalos with $R_{\rm max} > 5\times \epsilon$ are considered as well resolved in section\,\ref{sec:cvf}. 
The validity of this criterion is shown in subsection\,\ref{sec:densityprofile}. 
Finally, effects of the box size and the gravitational softening length on CVFs are investigated in subsection\,\ref{sec:cumulative} by taking into account intrinsic halo-to-halo scatter of the CVF. 
The simulations with $L = 5\,{\rm Mpc}/h$ in this appendix are the same simulations as shown in section\,\ref{sec:evolution} and details are given in subsection\,\ref{sec:mf}. 
In additional simulations in this appendix, we do not take the same samples of the initial fluctuations as in the simulations with $L = 5\,{\rm Mpc}/h$ in section\,\ref{sec:evolution}.
Let us stress that this appendix is not intended to check the rigorous convergence of simulations.

\subsection{Angular momenta}
\label{sec:angularmomenta} 
Our simulations in section\,\ref{sec:evolution} are performed with the relatively small box size $L = 5\,{\rm Mpc}/h$.
One may wonder if this may compromise the validity of our results completely.
It is suggested that the angular momentum of halos is possibly most sensitive to the simulation box size\,\cite{2006MNRAS.370..691P}.
This is because DM halos develop their angular momenta from the torque derived from large scale structure. 
Small simulation volumes may prevent DM halos from obtaining a correct amount of angular momentum. 
The change in angular momentum is reported to be mild, and specifically a $\sim 15\%$ systematic reduction when the simulation does not incorporate large scale ($\sim100\,{\rm Mpc}/h$) fluctuations.

In order to investigate this effect, we perform additional simulations with $L = 100\,{\rm Mpc}/h$ in the CDM model and measure the spin parameters of the simulated halos. 
The spin parameter is the dimensionless angular momentum defined as
\beq
\lambda = J/\sqrt{2}M_{\rm vir} r_{\rm vir} V_{\rm vir} \label{eq:bullockspin}
\eeq
in ref.\,\cite{Bullock2001}. 
Here, $J$, $M_{\rm vir}$, $r_{\rm vir}$, and $V_{\rm vir}$ are the angular momentum within the virial radius, the virial mass, the virial radius, and the circular velocity at the virial radius, respectively. 
In the additional simulations with $L = 100\,{\rm Mpc}/h$, the particle mass and the gravitational softening length are $m_{\rm part} \simeq 1.5\times 10^7 M_\odot/h$ and $\epsilon = 3000\,{\rm pc}/h$, respectively. 
We generate two realizations. 
The simulations with $L = 5\,{\rm Mpc}/h$ in the CDM model are the same as in section\,\ref{sec:evolution}.
Note that samples of the initial fluctuations are not identical between the simulations with $L = 5\,{\rm Mpc}/h$ and $L = 100\,{\rm Mpc}/h$.
The virial overdensity with respect to the critical density is set to be $96.8$ in the calculation of the spin parameter\,\cite{1997PThPh..97...49N}. 
The spin parameters are measured for halos whose masses are smaller than $10^{15} M_\odot/h$ and larger than $10^{10} M_\odot/h$.

\begin{figure}[tbp]
\centering 
\includegraphics[width=\hsize]{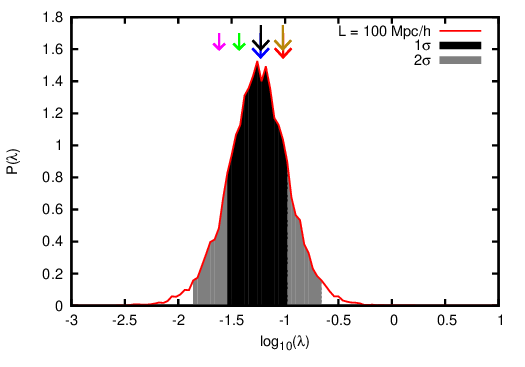}
\caption{\label{fig:spin} Distribution of the spin parameter in the additional simulations with $L=100\,{\rm Mpc}/h$ (two realizations) in the CDM model. 
The distribution $P(\lambda)$ is normalized as $\int P(\lambda) {\rm d} \log_{10} \lambda = 1$. 
The $1\sigma$ and $2\sigma$ regions are indicated by the black and the gray bands, respectively.
The vertical arrows show the spin parameters of our candidate halos of the MW and the Andromeda galaxy in the simulations with $L=5\,{\rm Mpc}/h$ (six realizations) investigated in subsection\,\ref{sec:cvf}.
The line colors are the same as those in figures\,\ref{fig:cvf} and\,\ref{fig:radis}.}
\end{figure}

It may be preferable to compare directly the distributions of the spin parameter between the simulations with $L=5\,{\rm Mpc}/h$ and $L=100\,{\rm Mpc}/h$.
However, there are sizable Poisson errors in the distribution of the spin parameter in the simulations with $L=5\,{\rm Mpc}/h$ (even with six realizations).
Instead, let us compare the distribution of the spin parameter in the simulations with $L=100\,{\rm Mpc}/h$ (two realizations) and the spin parameters of our candidate halos of the MW and the Andromeda galaxy in the simulations with $L=5\,{\rm Mpc}/h$ in figure\,\ref{fig:spin}.
The Poisson errors are negligible (sub \% level) and thus omitted in the figure.
Five out of the six halos have their spin parameters within $1\sigma$ region, while all the halos within $2\sigma$ region.
This implies that our candidate halos investigated in subsection\,\ref{sec:cvf} do not fail in capturing correct amounts of angular momentum.
We confirm this point in the next section by investigating their density profiles.

\subsection{Density profiles}
\label{sec:densityprofile}

\begin{figure}[tbp]
\centering 
\includegraphics[width=\hsize]{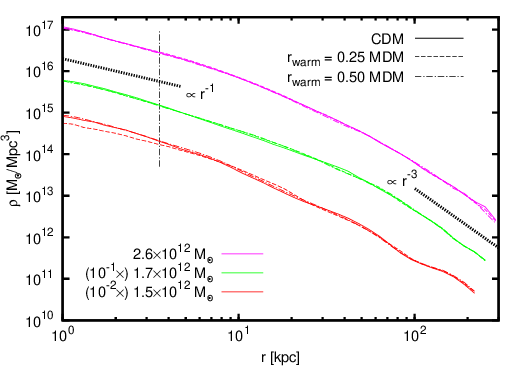}
\caption{\label{fig:profile} Comparison of density profiles. 
We show nine profiles of different halos in different DM models. 
The DM models are the CDM model (solid lines), the MDM models with $r_{\rm warm}=0.25$ (dashed lines) and $0.50$ (dash-dotted lines). 
The simulations are the same as in section\,\ref{sec:evolution}.
The halo masses are $\sim 2.6 \times 10^{12}M_\odot$ (magenta), $\sim 1.7\times 10^{12}M_\odot$ (green), and $1.5\times 10^{12}M_\odot$ (red). 
The line colors correspond to those in figures\,\ref{fig:cvf} and\,\ref{fig:radis}. 
For presentation, the profiles are shifted below for the halos with $\sim 1.7\times 10^{12}M_\odot$ and $1.5\times 10^{12}M_\odot$ by a factor of $10^{-1}$ and $10^{-2}$, respectively. 
The black dotted lines represent the inner and the outer slopes of the NFW profile. 
The black vertical dash-dotted line indicates the radius of $5\times \epsilon$, outside which the halos are expected to be well resolved. 
The difference between the profiles in the different DM models is small outside $5\times \epsilon$.}
\end{figure}

In the previous section, we investigated the spin parameters of our simulated halos in the CDM model.
We showed that all our candidate halos of the MW and the Andromeda galaxy are within $2\sigma$ region 
of the distribution of the spin parameter measured from the additional simulations with $L=100\,{\rm Mpc}/h$.
Since the spin parameters are expected to affect the density profiles of halos, let us examine their density profiles to confirm the conclusion in the previous subsection.
Figure\,\ref{fig:profile} 
displays the density profiles of three (different colors) of our candidate halos in the CDM model investigated in subsection\,\ref{sec:cvf}.
All the three profiles have an inner and an outer slope of $r^{-1}$ and $r^{-3}$ (black dotted lines), respectively, which are in good agreement with NFW profile.
This supports the previous implication that our candidate halos do not fail in capturing correct amounts of angular momenta.
This is because if they fail, their inner density profiles are expected to be steeper than NFW profile\,\cite{2004ApJ...604...18W}.

In figure\,\ref{fig:profile}, we show a radius of $5\times \epsilon \simeq 3.6\,{\rm kpc}$ by the black vertical dash-dotted line. 
The density profiles of our candidate halos in the CDM model are in good agreement with NFW profile outside $5\times \epsilon$.
In fact, the inner slope is consistent with $r^{-1}$ outside $\sim1\,{\rm kpc}\,(= 2 \times \epsilon)$.
Therefore, (sub)halos are expected to be well resolved if $R_{\rm max} > 5\times \epsilon$ in our simulations with $L=5\,{\rm Mpc}/h$.
We examine effects of the gravitational softening length further by comparing CVFs in high and low resolution simulations in the next section.

Before investigating CVFs, let us discuss the MW mass estimate in subsection\,\ref{sec:cvf}.
Reference\,\cite{Klypin2002} assumes the NFW profile to model the DM density profile.
We use their estimate to select the candidate halos of the MW in the MDM models, in which it is not clear if the DM particles follow the NFW profile.
This point should be addressed.
In figure\,\ref{fig:profile}, we plot the density profiles of corresponding halos in the MDM models with $r_{\rm warm}=0.25$ (dashed lines) and $0.50$ (dash-dotted lines).
Note that we take the identical samples of the initial fluctuations among the CDM and the MDM models in each realization except for the linear matter power spectra.
In each halo with different mass, the density profiles coincide with each other outside $5 \times \epsilon$.
The slopes of the density profiles at inner and outer regions coincide with those of the NFW profile.
Note that the DM mass density is only important in the radius larger than $\sim 5\,{\rm kpc}$, inside which the disk+bulge mass density dominates.
In addition, recent studies also imply that DM particles follow the NFW profile in WDM models ($k_{\rm J} = 34\text{--}60\,{\rm Mpc}^{-1}$)\,\cite{Lovell2014} and MDM models ($r_{\rm warm} = 0.1, 0.5, 1.0$ and $k_{\rm J} = 14, 32, 46\,{\rm Mpc}^{-1}$, respectively)\,\cite{2013MNRAS.428..882M}.
Therefore we can use the MW mass derived by assuming the NFW profile not only in the CDM model but also in the MDM models.

\subsection{Cumulative circular Velocity Functions}
\label{sec:cumulative}
\begin{figure}[tbp]
\centering 
\includegraphics[width=\hsize]{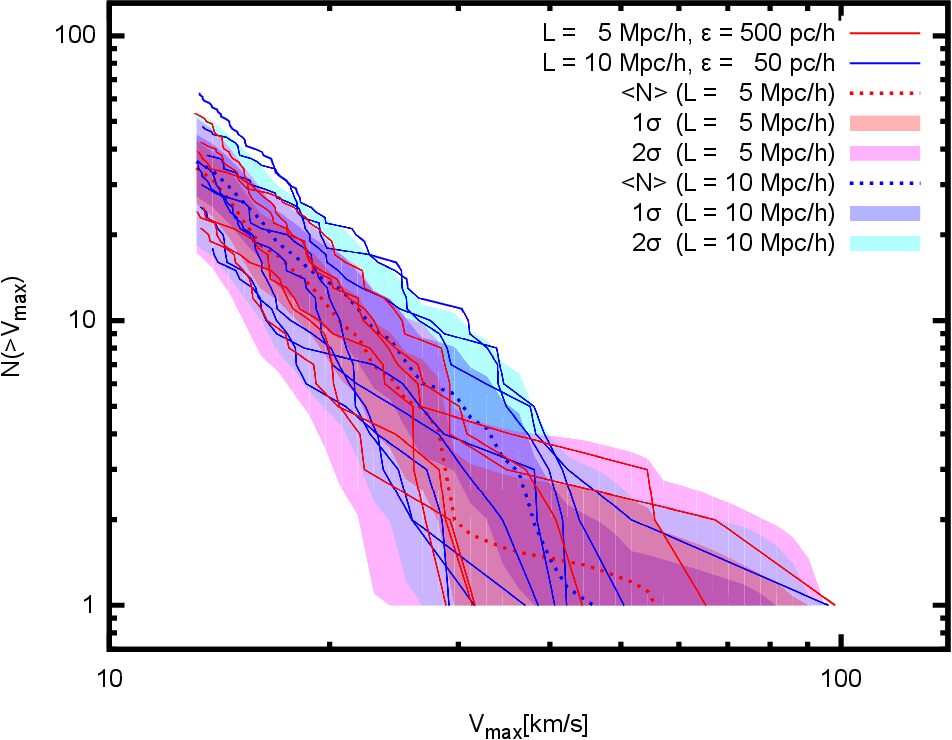}
\caption{\label{fig:compcvf} Comparison of the CVFs in the CDM model between the simulations with $L=5\,{\rm Mpc}/h$ and $\epsilon=500\,{\rm pc}/h$ (six realizations, red solid) and the additional simulation with $L=10\,{\rm Mpc}/h$ and $\epsilon = 50\,{\rm pc}/h$ (one realization, blue solid).  
All the CVFs are measured in the same manner in subsection\,\ref{sec:cvf}.
We plot $\langle N \rangle$'s (red dotted for the simulations with $L=5\,{\rm Mpc}/h$ and blue dotted for the additional simulation with $L=10\,{\rm Mpc}/h$) and $1\sigma$ (red for the simulations with $L=5\,{\rm Mpc}/h$ and blue for the additional simulation with $L=10\,{\rm Mpc}/h$) and $2\sigma$ (magenta for the simulations with $L=5\,{\rm Mpc}/h$ and cyan for the additional simulation with $L=10\,{\rm Mpc}/h$) regions of the estimated Negative Binomial distributions.}
\end{figure}

Finally, we compare the CVFs measured from the simulation with $L=5\,{\rm Mpc}/h$ and $\epsilon=500\,{\rm pc}/h$ in the CDM model to those from an additional simulation in the CDM model in figure\,\ref{fig:compcvf}.
The additional simulation has a larger box size $L=10\,{\rm Mpc}/h$, a larger particle mass $m_{\rm part} \simeq 5.7 \times 10^5 M_\odot/h$, and a smaller gravitational softening length $\epsilon = 50\,{\rm pc}/h$. 
Samples of the initial fluctuations in the additional simulation are not the same as in the simulations with $L=5\,{\rm Mpc}/h$ investigated in section\,\ref{sec:evolution}.
Let us note that the masses of the candidate halos are in the range discussed in subsection\,\ref{sec:cvf} both in the simulations with $L=5\,{\rm Mpc}/h$ and in the additional simulation with $L=10\,{\rm Mpc}/h$ but the masses themselves are different between them.
We restrict the comparison within the range of classical dwarfs. 
As discussed in subsection\,\ref{sec:cvf} and seen in figure 26 of ref.\,\cite{Springel2008}, $R_{\rm max}$ is several $\rm kpc$ for subhalos in the classical dwarf range. 
With the gravitational softening length of $\epsilon = 50\,{\rm pc}/h$ much smaller than $R_{\rm max}$, the additional simulation resolves subhalos sufficiently. 

While the simulations are initiated with different samples of the initial fluctuations, the CVFs are similar to each other within halo-to-halo scatter between the simulations with $L=5\,{\rm Mpc}/h$ and $\epsilon=500\,{\rm pc}/h$ and the additional simulation with $L=10\,{\rm Mpc}/h$ and $\epsilon=50\,{\rm pc}/h$.
We can confirm it qualitatively in the following way.
It is suggested that the cumulative number of subhalos at a given $V_{\rm max}$ follows the Negative Binomial distribution with the mean $\langle N \rangle$ and the variance $\sigma^2 = \langle N \rangle + (0.18 \langle N \rangle)^2$\,\cite{2010MNRAS.406..896B}. 
We estimate $\langle N \rangle$'s for a given $V_{\rm max}$ in the simulations with $L=5\,{\rm Mpc}/h$ and the additional simulation with $L=10\,{\rm Mpc}/h$ by maximizing the probabilities of obtaining the realized cumulative numbers of subhalos.
In figure\,\ref{fig:compcvf}, we present the estimated $\langle N \rangle$'s and $1\sigma$ and $2\sigma$ regions. 
The estimated distributions are close to each other. 
At the minimum of the maximum circular velocity $V_{\rm max}=13.16\,{\rm km/s}$, $\langle N \rangle = 34.3$ and $36.1$ for the simulations with $L=5\,{\rm Mpc}/h$ and the additional simulation with $L=10\,{\rm Mpc}/h$, respectively, and the difference is $\sim 5\%$.
Importantly, therefore, there is no implication of a significant reduction of the CVFs in the simulations with $L=5\,{\rm Mpc}/h$, which is expected to be seen if the gravitational softening length is not sufficiently small.
The subhalos appear to be well resolved in our simulations in subsection\,\ref{sec:cvf}.

\section{Difference between the halo maximum circular velocity and the peak of the rotation curve}
\label{sec:velconv}
In ref.\,\cite{Zavala2009}, they suggest that the ratio of the peak of the rotation curve of disk $V_{\rm max,\,d}$ to the maximum circular velocity of halo $V_{\rm max}$ takes the form
\beq
\frac{V_{\rm max,\,d}}{V_{\rm max}} = 1.04\left( 1-\frac{0.11f_{\rm disk} + 5 \times 10^{-4}}{\lambda^\prime}\right)^{-1} \,, \label{eq:diskcon}
\eeq
where $f_{\rm disk}$ is the ratio of the disk mass to the halo mass, and $\lambda^\prime = J\sqrt{|E|}/GM_{\rm vir}^{5/2}$ is the spin parameter defined in ref.\,\cite{Peebles1969}, and $E$ is the total energy of the halo within the virial mass. 
We adopt the choice in ref.\,\cite{Zavala2009} that $f_{\rm disk} = 0.03$. 
We consider the case that $\lambda^\prime$ takes the value from $0.014$ to $0.306$. 
This range is adopted for the following reason. 
In figure\,\ref{fig:spin}, halos have their spin parameters $\lambda$ (eq.\,(\ref{eq:bullockspin})) between $0.014$ and $0.219$ ($2\sigma$). 
The distribution of $\lambda$ does not change significantly in the MDM models. 
For the NFW profile halo, $\lambda = \lambda^\prime \sqrt{f(c)}$\,\cite{Maccio2007} where $c$ is the concentration parameter of the NFW profile and $f(c) = \frac{1}{2}c[(1+c)^2 - 1 - 2(1+c)\ln(1+c)]/[c-(1+c)\ln(1+c)]^2$\,\cite{Mo1998}. 
If the concentration parameter takes the value between 5 and 30\,\cite{Maccio2007}, 
the ratio $\lambda^\prime/\lambda$ takes the value between $1$ and $1.4$. 
We take the range of $\lambda^\prime$ as wide as possible, from $0.014\times1=0.014$ to $0.219\times1.4 = 0.306$. 
Then, the conversion factor (eq.\,(\ref{eq:diskcon})) takes the value between $1.05$ and $1.42$. 
The value is so close to unity that we ignore the difference between the peak of the rotation curve of disk and the maximum circular velocity of halo.

%% Figure %%
%\begin{figure}[tbp]
%\centering % \begin{center}/\end{center} takes some additional vertical space
%\includegraphics[width=.45\textwidth,trim=0 380 0 200,clip]{img1.pdf}
%\hfill
%\includegraphics[width=.45\textwidth,origin=c,angle=180]{img2.pdf}
% "\includegraphics" is very powerful; the graphicx package is already loaded
%\caption{\label{fig:i} Always give a caption.}
%\end{figure}

%% Table %%
%\begin{table}[tbp]
%\centering
%\begin{tabular}{|lr|c|}
%\hline
%x&y&x and y\\
%\hline
%a & b & a and b\\
%1 & 2 & 1 and 2\\
%$\alpha$ & $\beta$ & $\alpha$ and $\beta$\\
%\hline
%\end{tabular}
%\caption{\label{tab:i} We prefer to have borders around the tables.}
%\end{table}

\bibliographystyle{JHEP}
%when making bibtex file, you have to find articles from INSPIRE
\bibliography{ref}

\end{document}